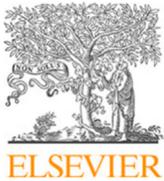
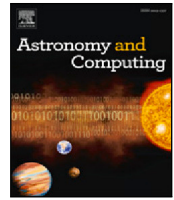

Full length paper

# Multivariate time-series forecasting of ASTRI-Horn monitoring data: A Normal Behavior Model☆

F. Incardona [a],*, A. Costa [a], F. Farsian [a], F. Franchina [a], G. Leto [a], E. Mastriani [a], K. Munari [a], G. Pareschi [b], S. Scuderi [c], S. Spinello [a], G. Tosti [d], for the ASTRI Project

[a] *INAF, Osservatorio Astrofisico di Catania, Catania, I-95123, Italy*
[b] *INAF, Osservatorio Astronomico di Brera, Merate, I-23807, Italy*
[c] *INAF, Istituto di Astrofisica Spaziale e Fisica Cosmica, Milano, I-20133, Italy*
[d] *Università di Perugia, Perugia, I-06123, Italy*

A B S T R A C T

This study presents a Normal Behavior Model (NBM) developed to forecast monitoring time-series data from the ASTRI-Horn Cherenkov telescope under normal operating conditions. The analysis focused on 15 physical variables acquired by the Telescope Control Unit between September 2022 and July 2024, representing sensor measurements from the Azimuth and Elevation motors. After data cleaning, resampling, feature selection, and correlation analysis, the dataset was segmented into fixed-length intervals, in which the first $I$ samples represented the input sequence provided to the model, while the forecast length, $T$, indicated the number of future time steps to be predicted. A sliding-window technique was then applied to increase the number of intervals. A Multi-Layer Perceptron (MLP) was trained to perform multivariate forecasting across all features simultaneously. Model performance was evaluated using the Mean Squared Error (MSE) and the Normalized Median Absolute Deviation (NMAD), and it was also benchmarked against a Long Short-Term Memory (LSTM) network. The MLP model demonstrated consistent results across different features and $I$–$T$ configurations, and matched the performance of the LSTM while converging faster. It achieved an MSE of $0.019 \pm 0.003$ and an NMAD of $0.032 \pm 0.009$ on the test set under its best configuration (4 hidden layers, 720 units per layer, and $I$–$T$ lengths of 300 samples each, corresponding to 5 h at 1-minute resolution). Extending the forecast horizon up to 6.5 h—the maximum allowed by this configuration—did not degrade performance, confirming the model's effectiveness in providing reliable hour-scale predictions. The proposed NBM provides a powerful tool for enabling early anomaly detection in online ASTRI-Horn monitoring time series, offering a basis for the future development of a prognostics and health management system that supports predictive maintenance.

## 1. Introduction

Modern astrophysical observatories are complex systems that integrate advanced mechanical, electronic, and software components, often combining telescopes into arrays to enhance observational capabilities. As the number of telescopes grows, the typical number of unexpected breakdowns increases proportionally, potentially reducing observation time and limiting the acquisition of scientific data. Equipment malfunctions, environmental factors, and aging infrastructure are common causes of operational downtime, often leading to substantial delays and unexpected maintenance costs.

Combining real-time monitoring, historical data analysis, and machine learning techniques offers a promising solution to address these challenges.

In this paper, we present a multivariate forecasting model designed to replicate the normal behavior of monitoring time series from the motor-related features of the ASTRI-Horn Cherenkov telescope. This model provides a powerful tool for enabling early fault detection in online ASTRI-Horn monitoring data by identifying anomalies as statistically significant deviations from normal behavior.

The proposed model could represent a valuable component within a broader Prognostics and Health Management (PHM) system (Lei et al., 2020; Zio, 2022). PHM comprises a set of methodologies designed to monitor the health of critical components, detect incipient faults, and estimate their Remaining Useful Life (RUL), thereby supporting informed and timely maintenance decisions. A central element of PHM is Condition-Based Monitoring (CBM), which refers to the process of continuously observing sensor data to assess equipment status in real time (Ali and Abdelhadi, 2022).

Within this framework, Predictive Maintenance (PdM) is the operational strategy that uses PHM outputs—such as anomaly indicators, diagnostics, and RUL estimates (Taşcı et al., 2023)—to anticipate failures and schedule maintenance before breakdowns occur.

Unlike traditional maintenance strategies—specifically, *preventive maintenance*, which relies on scheduled inspections and routine servicing to reduce the likelihood of breakdowns, and *corrective maintenance*, which focuses on repairing equipment after a failure has occurred—PdM aims to forecast potential failures before they escalate into critical damage, thus enabling early and targeted interventions (Molęda et al., 2023). This approach helps reduce maintenance costs and unplanned downtime, which, in the case of a telescope like ASTRI-Horn, has a direct impact in terms of increased observing time (Gambadoro et al., 2023). The proposed model, therefore, constitutes a first step toward enabling PdM capabilities for the ASTRI-Horn telescope (Incardona et al., 2024a), with potential applicability to other astronomical infrastructures.

*1.1. ASTRI-Horn*

ASTRI-Horn is a Cherenkov telescope designed to detect very high-energy gamma rays (Pareschi, 2016; Scuderi, 2018). It was developed as a prototype for the Italian-led ASTRI (Astrofisica con Specchi a Tecnologia Replicante Italiana) Project (Vercellone et al., 2013), a planned ensemble of nine dual-mirror telescopes to be deployed at the Observatorio del Teide in Tenerife, Spain. The array aims to study gamma-ray emissions from a wide range of astrophysical sources, including supernova remnants, pulsar wind nebulae, and active galactic nuclei (Vercellone et al., 2022; Scuderi et al., 2022).

Deployed in 2014 at the INAF "M.C. Fracastoro" station in Serra La Nave (1725 m a.s.l. Maccarone et al., 2013; Leto et al., 2023), Sicily, the ASTRI-Horn prototype represents the first implementation of dual-mirror Schwarzschild-Couder optics in ground-based gamma-ray astronomy (Giro et al., 2017). This optical design provides a wide field of view (approximately 10° (Rodeghiero et al., 2016)) with high angular resolution and minimal off-axis aberrations, essential for detecting faint, extended gamma-ray sources (Lombardi et al., 2020). The primary mirror comprises 18 hexagonal segments designed for high reflectivity and efficient photon collection (Canestrari et al., 2013). The secondary mirror, a monolithic structure with a diameter of 1.8 m, focuses the Cherenkov light on the focal plane (Catalano et al., 2018). The advanced telescope camera (Conforti et al., 2016), equipped with silicon photomultipliers (SiPMs), enables high-sensitivity measurements of atmospheric Cherenkov light generated by gamma-ray-induced particle cascades (Canestrari et al., 2019). This innovative design positions ASTRI-Horn as a crucial prototype for the Small-Sized Telescopes (SSTs) envisioned within the Cherenkov Telescope Array Observatory (CTAO (The CTA Consortium, 2018)).

As described by Sulich et al. (2024), the mechanical structure of the ASTRI-Horn prototype is built on an alt-azimuth mount, allowing extensive rotational movement with a range of ±270° in Azimuth and from 0° to +91° in Elevation. Its control system is based on a Beckhoff[1] programmable logic controller (PLC), which ensures precise real-time control, crucial for accurate tracking and pointing. The modular design of the PLC integrates high-speed processors and various I/O modules, providing fast response times while controlling the telescope's movements, positioning, power distribution, and safety protocols. Encoders provide position feedback on the Azimuth and Elevation axes, ensuring accurate tracking and alignment. The Azimuth axis is driven by two Beckhoff AM8062 servomotors in a master–slave configuration to reduce backlash and maintain precision, while the Elevation axis is powered by a single servomotor equipped with a brake and manual release for secure positioning when needed.

The Telescope Control Unit (TCU) oversees all aspects of telescope functionality, from movement to maintenance and safety (Tanci et al., 2016). The TCU generates pointing trajectories, coordinates subsystem operations (including the camera), and manages interlock signals and safety procedures in collaboration with a dedicated safety PLC. Communication between the TCU and high-level control software, such as the supervisory control and data acquisition (SCADA) system, is handled via OPC-UA[2] (Open Platform Communications – Unified Architecture). This secure, platform-independent protocol ensures seamless integration and efficient system operation (Bulgarelli et al., 2024; Russo et al., 2022; Conforti et al., 2022; Pastore et al., 2022).

To support its observational capabilities, ASTRI-Horn relies on an advanced software architecture built within an Internet of Things (IoT) framework, designed to manage and optimize the substantial data volumes generated during operations (Costa et al., 2020; Incardona et al., 2021a,b, 2022). The IoT-based infrastructure enables real-time monitoring and remote operations, enhancing scientific productivity and operational flexibility (Costa et al., 2021). Data collected include measurements from scientific instruments and from a network of sensors distributed throughout the telescope, which monitor parameters such as current, voltage, position, temperature, torque, motor and encoder status, and environmental conditions (e.g., humidity, wind speed, solar radiation (Leto et al., 2014)).

Sensor data not only support the correction of systematic measurement errors but also serve as input for the continuous assessment of system conditions through CBM. In this framework, the timely detection of abnormal patterns in time-series signals is crucial to the effectiveness of PdM interventions.

*1.2. Anomaly detection in time series with machine learning*

Time Series Anomaly Detection (TSAD) is the process of identifying data points or patterns that significantly deviate from expected trends over time (Sgueglia et al., 2022). Traditional approaches often rely on predefined thresholds or rule-based strategies, but the increasing volume and complexity of data have driven a shift toward data-driven methods capable of capturing the underlying dynamics of a time series, thereby achieving more accurate anomaly detection compared to conventional techniques (Blázquez-García et al., 2021; Correia et al., 2024).

Machine learning (ML), broadly defined as the study of algorithms and statistical models that enable computer systems to learn and make predictions or decisions without being explicitly programmed (Pugliese et al., 2021), can be divided into four principal paradigms. In *supervised learning*, each training instance is associated with a known label or outcome, and the model is trained to predict those labels in new, unseen data (Jia et al., 2019; Yokkampon et al., 2021). In *unsupervised learning*, no labels are provided, and the goal is to discover hidden structures within the data, such as clusters or low-dimensional representations (Ngo Bibinbe et al., 2022). *Semi-supervised learning* bridges these two approaches by combining labeled and unlabeled data to improve predictive performance when only a small portion of the data has labels (van Engelen and Hoos, 2020). Finally, *reinforcement learning* focuses on an agent that interacts with an environment to maximize a cumulative reward signal—an approach that can be adapted for anomaly detection if atypical states are associated with lower rewards (Arshad et al., 2022).

---

[1] https://www.beckhoff.com/

[2] https://opcfoundation.org/

Across all these paradigms, *deep learning* techniques have emerged as powerful tools for learning complex mappings from data, particularly when dealing with high-dimensional, unstructured, or time-dependent inputs. Architectures such as Recurrent Neural Networks (RNNs), Convolutional Neural Networks (CNNs), *autoencoders*, and *transformers* are especially well-suited for modeling large-scale or rapidly evolving time-series datasets (Li and Jung, 2023; Zamanzadeh Darban et al., 2024).

A frequently employed strategy in TSAD is to build a Normal Behavior Model (NBM) that approximates the typical states or trends within a dataset (Chesterman et al., 2023). This model then acts as a reference against which new observations are compared. Once a model is trained on normal data, deviations from its learned representation indicate potential anomalies (Katragadda et al., 2020). This approach is especially valuable when anomalies are rare or difficult to characterize with explicit rules, allowing the model to adapt to wide-ranging operational conditions (Xiao et al., 2022).

### 1.3. Strategy for early anomaly detection in ASTRI-Horn monitoring time series

Since beginning operations, the ASTRI-Horn prototype has experienced three significant issues that forced the suspension of telescope activities. In May 2019, water infiltration caused gear oxidation, seizing the Azimuth master motor. In early November 2021, gradual wear on the Azimuth relative encoder tape led to increasing errors; operations were halted until the encoder was replaced in April 2022. Upon inspection during the replacement, multiple points of damage were found on the tape. Finally, in February 2022, condensation-induced oxidation damaged the brake of the Elevation motor.

The findings of Incardona et al. (2024b) reveal insufficient evidence of gradual degradation in the time series associated with these three faults. Furthermore, the limited number of such incidents complicates the use of standard supervised learning algorithms for anomaly detection (Costa et al., 2024). However, a substantial dataset of normal operational data, available since September 2022, provides a robust basis for implementing an NBM built on a deep learning architecture (Section 1.2), specifically tailored to the telescope's time series.

The NBM is trained exclusively on time-series data reflecting Normal Operating Conditions (NOCs), which correspond to healthy operational states free from anomalies, degradations, or failures. During training, the model learns the expected behavioral patterns for each feature. In real-time operation, the NBM is intended to support continuous monitoring of each target feature by comparing measured values with model-generated predictions. Substantial deviations from expected behavior can serve as indicators for triggering alarms, pointing to potential anomalies or early signs of component degradation. To quantify such deviations, the residuals can be computed as the difference between observed and predicted values for each feature at a given timestamp.

Specifically, for a feature $f$ and data point $i$, the residuals can be defined as:

$$r_{f,i} = \text{Observed}_{f,i} - \text{Predicted}_{f,i}. \tag{1}$$

With multiple features, these residuals can be aggregated into an *anomaly score*, $a_i$, for point $i$:

$$a_i = \sqrt{\sum_{f=1}^{F} r_{f,i}^2}, \tag{2}$$

where $F$ is the total number of features.

The anomaly score can then be normalized to permit consistent thresholding across various features and operating conditions (Aggarwal, 2017). Any score exceeding a predefined threshold will be flagged as an anomaly, allowing operators to investigate and address potential issues. To enhance interpretability, feature heatmaps are expected to offer an intuitive view of deviations across multiple features and time periods (Tong et al., 2023).

In addition to static thresholding, residuals can be further processed using, for example, the CUmulative SUM (CUSUM) method, which incrementally accumulates small deviations over time, enabling the identification of subtle or slowly evolving anomalies that may not immediately breach static thresholds (Page, 1954; Xie et al., 2021). This property makes CUSUM particularly well-suited for scenarios involving mechanical wear or progressive misalignments (Jawad and Jaber, 2021).

## 2. Historical monitoring time series of ASTRI-Horn

The operational history of the ASTRI-Horn prototype is recorded in a database of 479 tables, each corresponding to a time series of a monitored property collected by a sensor. Each table is associated with a specific *assembly*, representing a physical device under observation.

For this analysis, data ranging from September 6, 2022, to July 8, 2024, were selected to ensure consistent availability across the chosen features. Furthermore, the study was narrowed to the TCU-related properties (see Section 1), focusing specifically on the telescope's motors, which constitute critical and high-cost mechanical components. Despite this focus, data retrieval and analysis remained challenging. Properties containing no data, exhibiting fixed values throughout, or deemed non-essential based on documentation were removed. Categorical variables were also excluded. As a result, a subset of 15 features was selected, which are detailed in Table 1.

**Table 1**
Selected TCU features analyzed in this study.

| Feature | Description | Units |
| --- | --- | --- |
| AZACTPOS | Current position along the Azimuth axis. | ° |
| AZACTVEL | Current velocity along the Azimuth axis. | °s$^{-1}$ |
| AZMASTERCURRENT | Current consumption of the Azimuth master motor. | A |
| AZMASTERTEMP | Temperature of the Azimuth master motor. | °C |
| AZMASTERTORQUE | Torque of the Azimuth master motor. | N m |
| AZMASTERVOLTAGE | Voltage supplied to the Azimuth master motor. | V |
| AZSLAVECURRENT | Current consumption of the Azimuth slave motor. | A |
| AZSLAVETEMP | Temperature of the Azimuth slave motor. | °C |
| AZSLAVETORQUE | Torque of the Azimuth slave motor. | N m |
| AZSLAVEVOLTAGE | Voltage supplied to the Azimuth slave motor. | V |
| ELACTPOS | Current position along the Elevation axis. | ° |
| ELACTVEL | Current velocity along the Elevation axis. | °s$^{-1}$ |
| ELMOTORCURRENT | Current consumption of the Elevation motor. | A |
| ELMOTORTEMP | Temperature of the Elevation motor. | °C |
| ELMOTORTORQUE | Torque of the Elevation motor. | N m |

### 2.1. Resampling

Most time series were sampled at approximately 15-second intervals, although some exhibited variable sampling frequencies. To ensure consistency, the data were resampled to a uniform 1-minute interval by averaging the values within each minute (typically about four samples). This choice also reduces the number of samples to a more computationally manageable level while still providing sufficient temporal resolution for a forecasting task targeting hour-scale trends.

### 2.2. Active motor intervals

To maximize the information content of the time series and focus exclusively on relevant data, the raw telescope motor data were restricted to intervals when at least one of the two motors was active. These active motor intervals were identified by analyzing the AZMOTORSTATUS and ELMOTORSTATUS features, selecting only periods in which either indicator exceeded zero. This selection process yielded a total of 325 active intervals, whose amplitude distribution—measured in hours—is shown in Fig. 1.

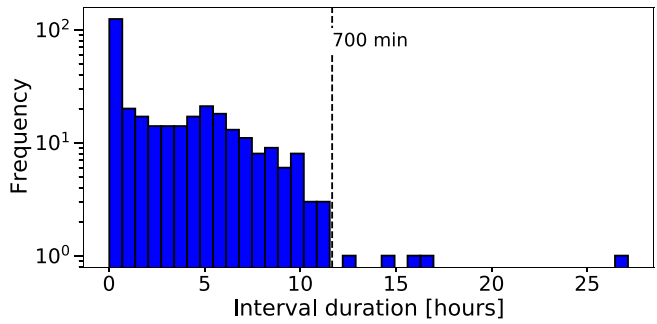

**Fig. 1.** Histogram of the time durations (in hours) of the active motor intervals. The *x*-axis represents interval durations, and the *y*-axis (log scale) indicates the frequency. The black dashed vertical line marks the threshold (in minutes) below which each duration range contains more than one interval.

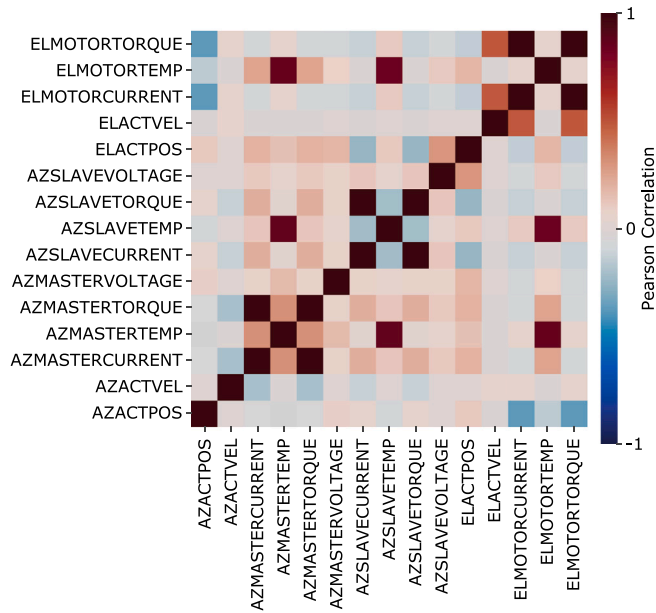

**Fig. 2.** Correlation heatmap of the selected TCU features. The color scale spans from −1 (strong negative correlation, shown in purple) to +1 (strong positive correlation, shown in yellow), with 0 (green) indicating no correlation.

The histogram reveals that most intervals are relatively short, ranging between 0 and approximately 40 minutes, with many others spanning from 40 minutes to 10 h. A long tail extends up to about 27 h. The mean duration is approximately 3.3 h, while the median is about 2.15 h.

### 2.3. Correlation studies

We analyzed the correlations among the selected features.

The heatmap shown in Fig. 2 visualizes Pearson coefficients,[3] where values near +1 or −1 indicate strong positive or negative correlations, respectively. While most features appear uncorrelated, certain regions in the heatmap reveal clusters of highly correlated features. Table 2 summarizes all correlations above 1 sigma ($|r| \geq 0.6827$).

The analysis revealed a correlation above 3 sigma between torque and current for both the Azimuth and Elevation motors. This is consistent with the fact that, in electric motors, torque is directly proportional

---

[3] Defined as the covariance of two variables divided by the product of their standard deviations (Johnson and Wichern, 1988).

**Table 2**
Correlation results for the selected TCU features.

| Feature 1 | Feature 2 | Pearson Coefficient |
|---|---|---|
| AZSLAVETORQUE | AZSLAVECURRENT | 0.999 |
| AZMASTERTORQUE | AZMASTERCURRENT | 0.999 |
| ELMOTORTORQUE | ELMOTORCURRENT | 0.998 |
| AZSLAVETEMP | AZMASTERTEMP | 0.869 |
| ELMOTORTEMP | AZMASTERTEMP | 0.867 |
| ELMOTORTEMP | AZSLAVETEMP | 0.848 |

to the current, according to:

$$\tau = k_T \cdot \mathcal{I}, \qquad (3)$$

where $\tau$ is the torque, $\mathcal{I}$ is the current, and $k_T$ is the torque constant, a characteristic of the motor (Hughes and Drury, 2019). Consequently, we removed the three torque features from the dataset to reduce redundancy.

### 2.4. Removal of outliers

We cleaned the dataset of clearly erroneous data—only a few instances across the entire dataset—consisting of physically impossible readings or obvious sensor misreadings, which do not reflect any mechanical degradation or component faults. These outlier values were replaced using a combination of backward filling (replacing invalid entries with the next valid observation) and forward filling (substituting with the most recent valid value).

### 2.5. Final dataset organization

The final dataset consisted of 64,669 rows, divided into 325 intervals, and 12 columns representing the features. It was structured using a multi-level index[4] combining interval and timestamp information.

Fig. 3 shows the time series for the 12 selected TCU motor-related features after preprocessing. The curves are not continuous because only selected intervals were used. Clear seasonal patterns are visible in AZMASTERTEMP, AZSLAVETEMP, and ELMOTORTEMP.

## 3. Methods: Multivariate forecasting

The development of the NBM was driven by the need to address the challenges outlined in Section 1.3. By training on TCU data collected under NOCs, the model learns behavioral patterns that are intended to enable the detection of anomalies as deviations from these learned trends.

Because of the interconnected nature of the features in the telescope's time series, the model must capture both temporal dynamics and interdependencies among variables. Hence, a multivariate forecasting approach was adopted, allowing the prediction of multiple time-series variables simultaneously rather than one variable in isolation.

### 3.1. Multi-layer perceptron

We employed a feedforward artificial neural network (ANN), specifically a Multi-Layer Perceptron (MLP).

MLPs learn mappings between inputs and outputs via multiple fully connected layers, each containing a number of neurons with activation functions that introduce non-linearity. Neuron connections are associated with weights, and during the forward pass, input data propagate through the network to compute output predictions based on the current weights and activation functions. Training is performed through backpropagation (Li, 2024), an algorithm that computes the gradient of the loss function (see Section 4.2) with respect to each weight, quantifying its contribution to the overall error. These gradients are then used during the backward pass to iteratively update the weights, allowing the model to minimize prediction errors and to capture complex interactions in the data (Lazcano et al., 2024).

---

[4] https://pandas.pydata.org/docs/reference/api/pandas.MultiIndex.html

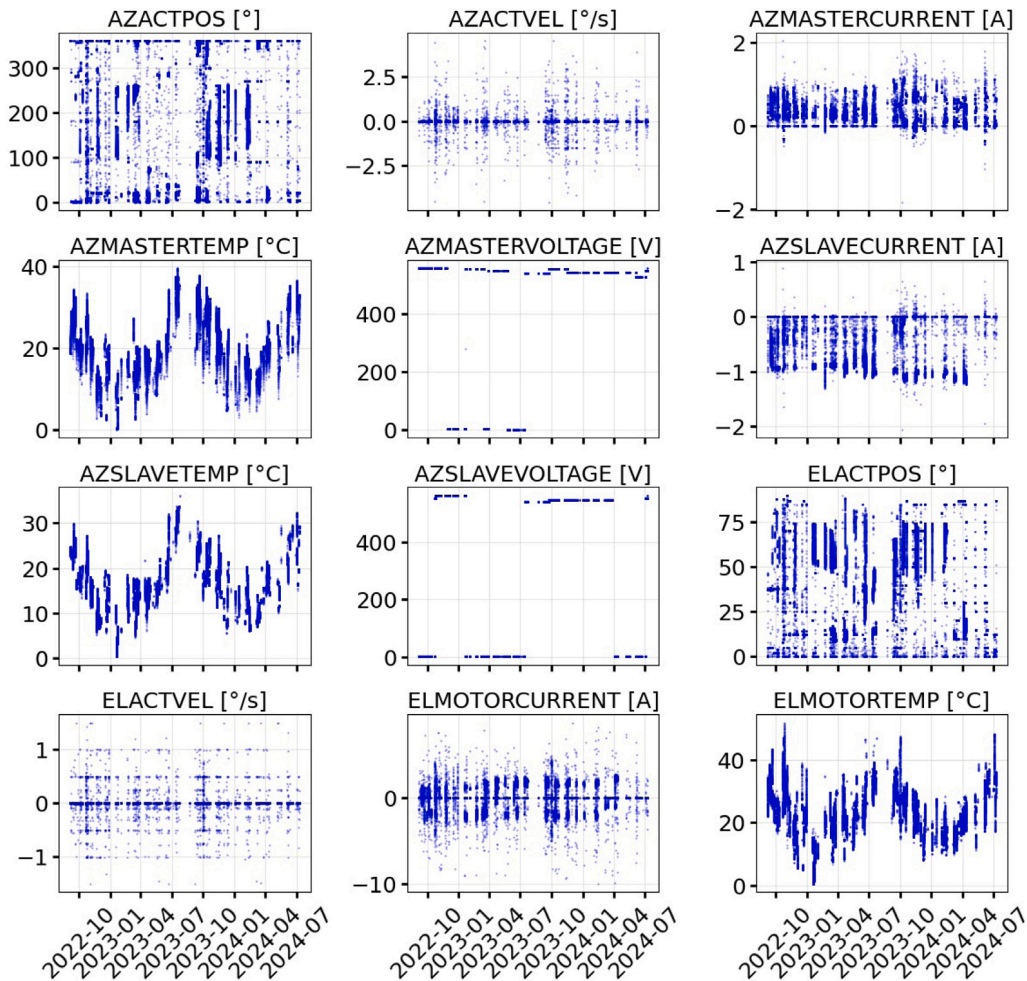

**Fig. 3.** The 12 selected TCU motor-related features after preprocessing. Each panel represents a different feature, with the *x*-axis indicating timestamps and the *y*-axis the feature values. The discontinuities correspond to selected active motor intervals.

The MLP includes an input layer followed by $L$ fully connected hidden layers, each containing $n$ neurons—the fundamental computational units of the network—and an output layer producing the final predictions.

We implemented this architecture in `PyTorch` (Paszke et al., 2019), using ReLU activation functions (Dubey et al., 2022) and the Adam optimizer to perform efficient gradient descent during training (Kingma and Ba, 2014). The architecture is shown in Fig. 4, while the composition of the input and output layers is detailed in Section 3.2.

### 3.2. Input and target data

From each active motor interval (Section 2.5), we extract chunks of fixed length $W$. For each test configuration (see Section 4), a specific value of $W$ is selected. Active motor intervals shorter than $W$ are discarded, while those exactly $W$ samples long are retained as a single chunk. Longer intervals are processed using the sliding-window technique described in Section 3.3 to extract multiple chunks of length $W$.

Within each chunk, the first $I$ time steps serve as input samples for the model, while the subsequent $T$ constitute the target samples to be predicted, ensuring:

$$W = I + T. \tag{4}$$

This configuration is instrumental in teaching the model to predict future values from the given input sequence.

All features are predicted simultaneously, preserving the multivariate nature of the forecast. Specifically, the input layer has dimension $I \times F$, where $I$ is the previously defined number of input samples and $F$ is the number of features. The output layer has dimension $T \times F$, corresponding to $T$ predicted time steps for all features. The input and output layers are illustrated in Fig. 4.

### 3.3. Sliding window

The sliding-window technique increases the amount of training data by systematically shifting a smaller window across a longer interval, thereby generating multiple fixed-length segments (Khandelwal et al., 2020). A schematic illustration is provided in Fig. 5.

In this study, we shift a $W$-length window across each active motor interval longer than $W$, producing multiple sub-intervals. The window advances one time step at a time, capturing similar but temporally offset input-target segments. Consequently, for each active motor interval of $N_i > W$ samples, $N_i - W + 1$ sequences are extracted, thereby increasing the dataset size. The total number of chunks $C_W$ for a given test configuration with a specific value of $W$, can then be computed as:

$$C_W = \sum_{i=1}^{325} f_W(N_i), \tag{5}$$

with:

$$f_W(N_i) = \begin{cases} 0 & \text{if } N_i < W \\ N_i - W + 1 & \text{if } N_i \geq W \end{cases}. \tag{6}$$

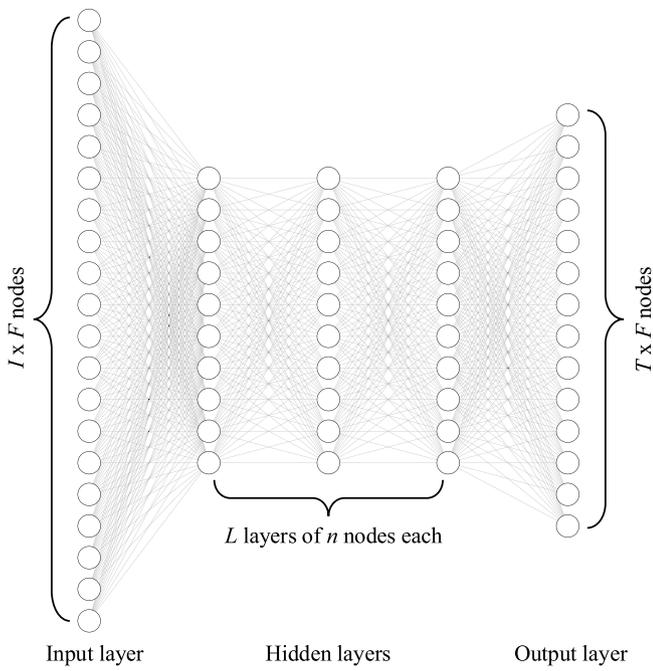

**Fig. 4.** Schematic architecture of the MLP used in this study. The network takes as input an $I \times F$ vector—where $I$ is the number of input samples and $F$ the number of features—and outputs a $T \times F$ prediction vector, corresponding to $T$ future steps for each feature. The hidden body of the network consists of $L$ fully connected layers, each with $n$ nodes.

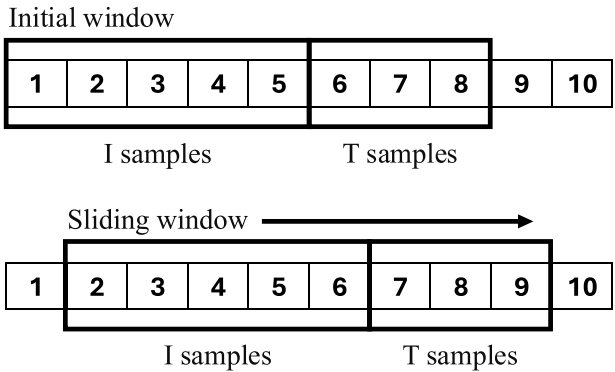

**Fig. 5.** Illustration of the sliding-window technique used to expand the dataset. As an example, the top row represents an initial window of length $W = 8$ samples over an interval of $N = 10$ samples, with its input ($I = 5$) and target samples ($T = 3$). The bottom row shows how the window slides across the time-series data, maintaining input and target samples for each step.

Since the frequency of longer intervals drops to isolated occurrences beyond 700 min (see Fig. 1), we set $W = 700$ as a practical upper limit for its selection in the test configurations. Larger values of $W$ would result in the exclusion of a substantial portion of available active motor intervals, and the application of the sliding-window technique would introduce redundancy without providing meaningful additional training examples.

### 3.4. Data preparation

To prepare the dataset for training, a few additional pre-processing steps were applied.

Each feature was normalized to ensure that all of them were on a comparable scale, mitigating the risk of certain features dominating the training process. We applied the *z-score* normalization method (Lima and Alves de Souza, 2023):

$$y_i = \frac{y'_i - \mu}{\sigma}, \tag{7}$$

where $y'_i$ is the original data point, and $\mu$ and $\sigma$ are the mean and standard deviation of the feature, respectively. This transformation centers the data around zero with a standard deviation of one.

Missing values were handled through interpolation, preserving temporal continuity.

The data were then partitioned into training, validation, and test sets in proportions of about 80%, 10%, and 10%, respectively. Clearly, to maintain the temporal structure of the sequences, the partitioning was performed on the basis of $W$-length chunks rather than on individual data points. Furthermore, the chunks assigned to each set were randomly sampled from the entire dataset to mitigate the influence of long-term patterns (e.g., seasonal effects). This approach preserves temporal correlations within each window while breaking the continuity of slower-varying trends across the splits.

Finally, input and output samples were converted into `PyTorch` tensors.

## 4. Model training, evaluation, and comparison

As explained in Section 3.2, the input time-series data were divided into $W$-length segments, which were internally split into $I$ input and $T$ target samples. We tested the model over different combinations of $W$, $I$, and $T$ to determine the optimal configuration. For each tested configuration, we tuned a subset of the hyperparameters, as discussed in Section 4.1.

### 4.1. Hyperparameter selection and tuning

Training was performed using a fixed batch size of 256, which specifies the number of elements processed in parallel during each forward pass (Section 3.1).

The learning rate, which controls the step size of weight updates at each iteration by acting as a multiplier of the gradient, was set to 0.001. Although other values (0.01 and 0.1) were tested as well, this value provided the best balance between convergence speed and stability.

To optimize the model architecture described in Section 3.1, we performed a grid search over a predefined set of values for the number of hidden layers, $L$, and units per layer, $n$. The tested values for $n$ were 100, 360, 720, and 1440, while $L$ ranged from 1 to 10. Each layer contained the same number of neurons. Architectures with a low number of neurons per layer ($n < 100$) lacked the complexity necessary to achieve a low validation loss, whereas those with a high number ($n > 1440$) were prone to overfitting and incurred significantly higher computational costs. These tested ranges trade off model complexity and generalization performance.

## 4.2. Evaluation metrics

We used the Mean Squared Error (MSE)[5] as both the training loss function and the evaluation metric on the training, validation, and test sets:

$$\text{MSE} = \frac{1}{T} \sum_{i=1}^{T} (y_i - \hat{y}_i)^2, \tag{8}$$

where $y_i$ and $\hat{y}_i$ are the true and predicted values, respectively, and $T$ is the number of target samples.

MSE is one of the most widely adopted loss functions in anomaly detection (Boggia et al., 2025), and it is not susceptible to instability when dealing with near-zero values.

To evaluate the performance of the final model (Section 5.2), in addition to MSE, we computed the Normalized Median Absolute Deviation (NMAD) of the residuals:

$$\text{NMAD} = 1.4826 \times \text{median}_{i=1,\ldots,T} \left( \left| (y_i - \hat{y}_i) - \text{median}_{j=1,\ldots,T} (y_j - \hat{y}_j) \right| \right). \tag{9}$$

Unlike MSE, which is dominated by large deviations due to its quadratic form, NMAD provides an estimate of the typical dispersion of the residuals around their median (Luo et al., 2024).

## 4.3. Test cases comparison

We explored the MLP forecasting capabilities across various configurations of $W$, $I$, and $T$ time steps, as well as different architectures. For each configuration, we trained up to 500 epochs, with each epoch representing a complete processing of the entire training dataset by the model, and applied early stopping if the validation loss did not improve after 50 consecutive epochs.

To benchmark the performance of the MLP model, we also implemented a more complex architecture: the Long Short-Term Memory (LSTM) network.

LSTMs are a specialized type of RNN designed to capture long-term dependencies in sequential data while mitigating the vanishing gradient problem (Lindemann et al., 2021). Unlike MLPs, which process the entire input as a fixed-length vector without accounting for temporal order, LSTMs process sequences one element at a time, retaining information from earlier elements in the same input sequence through gated memory mechanisms. This enables them to model dependencies across time steps and better capture temporal patterns (Abdulrahim et al., 2025).

The LSTM architecture was also implemented using `PyTorch` and optimized over a set of hyperparameters, including the number of hidden units per layer, $\eta$, and the number of stacked LSTM layers, $\ell$. Specifically, we tested $\eta = 100$, 360, and 720, and varied $\ell$ from 1 to 3.

The optimal hyperparameters for each $W$ and $I$–$T$ combination are listed in Table 3. They were selected based on the lowest loss score achieved on the validation set during training (hereafter referred to as validation loss).

The results reveal that, when the number of input/target samples is extremely limited ($W = 6$), MLP and LSTM exhibit similarly poor performance. For $W = 60$, however, the LSTM consistently outperforms the MLP, often by a substantial margin, likely due to its capacity to exploit internal memory. As $W$ increases to 360, the performance gap narrows again, and in the specific case of $I = 300$ and $T = 60$, the MLP marginally surpasses the LSTM, suggesting that the MLP might require more data to reach the LSTM's performance level.

---

[5] https://scikit-learn.org/1.5/modules/model_evaluation.html#mean-squared-error

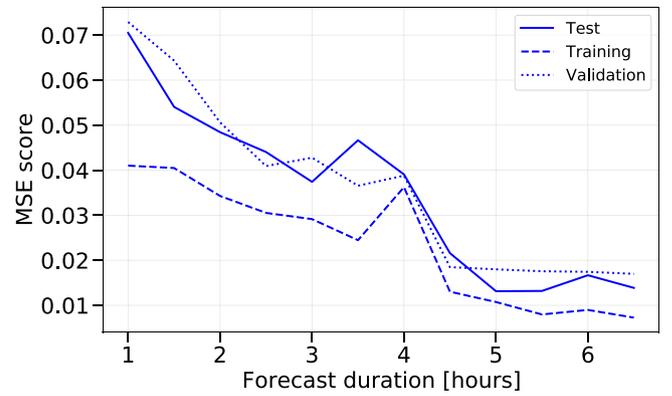

**Fig. 6.** MSE score of the MLP models as a function of forecast duration, with the $x$-axis indicating the forecast horizon in hours. The plot shows the MSE for the test (solid line), training (dashed line), and validation (dotted line) datasets. The forecast duration ranges from 1 to 6.5 h, corresponding to target samples ranging from 60 to 390, respectively.

In most cases, the best architectures rely on 100 or 360 internal units per layer, even when trained on small datasets, although the optimal number of layers may vary. MLPs generally require deeper architectures to remain competitive, whereas LSTMs tend to capture temporal patterns more efficiently with fewer layers. Nonetheless, from a computational standpoint, MLPs converge approximately ten times faster than LSTMs, making them a particularly appealing choice for larger values of $W$, where their performance becomes comparable.

## 4.4. Increasing the number of target samples

Given their faster convergence and competitive performance, MLPs were selected for the subsequent analysis to assess the models' performance on larger prediction horizons.

We fixed $I = 300$ and progressively increased the number of target samples, $T$, from 60 to 390 in increments of 30. The upper limit $T = 390$ was constrained by the choice of a maximum window size $W = 700$, as discussed in Section 3.3. For each of the incremental steps, we searched for the best architecture. Fig. 6 highlights the behavior of the MSE score across test, training, and validation sets as a function of $T$, expressed in hours. The conversion is straightforward given the 1-minute resampling (Section 2.1).

The results show a clear trend of decreasing MSE as $T$ increases, with a significant performance improvement up to approximately 5 h ($T = 300$), plateauing beyond this point. This suggests that the model reaches a regime in which additional target samples do not significantly enhance generalization, indicating a saturation in learning efficiency. In this region, the best-performing architectures tend to have fewer layers (3 or 4) but maintain a high number of units per layer (720), as shown in Table 4.

The findings of these tests indicate that the model effectively captures multi-hour trends, achieving the best result of MSE = 0.0131 for $T = 300$ in the test set. However, although the validation and test curves tend to converge at larger $T$ values, the training loss remains systematically lower than both the validation and test curves, pointing to some degree of overfitting despite the overall performance gains.

## 4.5. Loss curve

To delve deeper into the potential overfitting issue, we show in Fig. 7 the training and validation loss curves for the MLP model that achieved the best MSE score on the test set (4 hidden layers, 720 neurons per layer, $I = T = 300$). These curves provide a direct measure of the model's parameter adaptation during training and its resulting

Table 3

Summary of the test cases analyzed with the MLP and LSTM models, trained for 500 epochs. The first three columns report the number of input samples ($I$), target samples ($T$), and the total window size ($W$), respectively. For each model type, the best architecture is reported: in terms of number of units per layer ($n$) and layers ($L$) for the MLP, and number of hidden units ($\eta$) and stacked layers ($\ell$) for the LSTM. The corresponding best MSE scores on the training, validation, and test sets are also provided.

| Number of Samples | | | MLP | | | | | LSTM | | | | |
|---|---|---|---|---|---|---|---|---|---|---|---|---|
| | | | Best Architecture | | MSE | | | Best Architecture | | MSE | | |
| $I$ | $T$ | $W$ | $n$ | $L$ | Training | Validation | Test | $\eta$ | $\ell$ | Training | Validation | Test |
| 5 | 1 | 6 | 360 | 3 | 0.0739 | 0.1543 | 0.1833 | 360 | 1 | 0.0669 | 0.1592 | 0.1864 |
| 4 | 2 | 6 | 360 | 2 | 0.1024 | 0.1729 | 0.1748 | 100 | 2 | 0.0877 | 0.1591 | 0.1645 |
| 3 | 3 | 6 | 100 | 2 | 0.1385 | 0.1489 | 0.1555 | 100 | 3 | 0.1006 | 0.1580 | 0.1691 |
| 50 | 10 | 60 | 720 | 7 | 0.0707 | 0.1429 | 0.1495 | 360 | 3 | 0.0284 | 0.0560 | 0.0582 |
| 40 | 20 | 60 | 720 | 9 | 0.0908 | 0.1396 | 0.1459 | 360 | 2 | 0.0246 | 0.0547 | 0.0578 |
| 30 | 30 | 60 | 360 | 10 | 0.1009 | 0.1399 | 0.1446 | 360 | 2 | 0.0213 | 0.0546 | 0.0607 |
| 300 | 60 | 360 | 360 | 7 | 0.0410 | 0.0729 | 0.0705 | 100 | 2 | 0.0669 | 0.0746 | 0.0746 |
| 240 | 120 | 360 | 360 | 10 | 0.0519 | 0.0614 | 0.0697 | 360 | 2 | 0.0437 | 0.0571 | 0.0653 |
| 180 | 180 | 360 | 360 | 8 | 0.0471 | 0.0677 | 0.0718 | 360 | 2 | 0.0425 | 0.0639 | 0.0698 |

Table 4

Performance of the MLP models for a fixed input size ($I = 300$) and increasing target sizes ($T$), trained for 500 epochs. The table presents the number of target ($T$) and total ($W$) samples, along with the best architecture found in terms of the number of units per layer ($n$) and the number of layers ($L$). In each case, the best MSE scores for the training, validation, and test sets are reported. The configuration with the lowest MSE score on the test set is highlighted in bold.

| Number of Samples | | Best Architecture | | MSE | | |
|---|---|---|---|---|---|---|
| $T$ | $W$ | $n$ | $L$ | Training | Validation | Test |
| 60 | 360 | 360 | 7 | 0.0410 | 0.0729 | 0.0705 |
| 90 | 390 | 360 | 5 | 0.0405 | 0.0643 | 0.0541 |
| 120 | 420 | 720 | 9 | 0.0343 | 0.0506 | 0.0484 |
| 150 | 450 | 360 | 6 | 0.0305 | 0.0409 | 0.0440 |
| 180 | 480 | 1440 | 7 | 0.0291 | 0.0428 | 0.0374 |
| 210 | 510 | 360 | 5 | 0.0245 | 0.0366 | 0.0466 |
| 240 | 540 | 360 | 5 | 0.0362 | 0.0388 | 0.0391 |
| 270 | 570 | 1440 | 3 | 0.0130 | 0.0185 | 0.0216 |
| **300** | **600** | **720** | **4** | **0.0107** | **0.0180** | **0.0131** |
| 330 | 630 | 720 | 3 | 0.0080 | 0.0176 | 0.0132 |
| 360 | 660 | 720 | 3 | 0.0090 | 0.0174 | 0.0167 |
| 390 | 690 | 720 | 3 | 0.0073 | 0.0170 | 0.0139 |

performance on the validation set, thereby illustrating how effectively the network generalizes to unseen data. The figure shows the mean loss curves over 10 repetitions of the training process, with shaded regions indicating one standard deviation. The average training duration was approximately 34 s using the Metal Performance Shaders backend[6] for GPU acceleration on Apple M3 hardware.

The plotted loss curves show that both training and validation losses decrease over the epochs, demonstrating effective learning and convergence of the model. The relatively small standard deviation across repetitions underscores the stability and robustness of the training process. The spikes in the loss curves do not significantly affect the overall trend and likely reflect transient variability, possibly due to noise in the input data or sensitivity to certain features. Finally, the slight divergence between the training and validation losses indicates potential overfitting, as the model begins to memorize the training data rather than learning patterns that generalize to new, unseen data. To mitigate this, training must be stopped at 112 epochs, the point at which the loss curves begin to diverge. Continuing beyond this point would not yield better generalization and, moreover, would not lead to significant performance improvements. Accordingly, we did not introduce dropout (Salehin and Kang, 2023) or include any regularization term in the loss function (Bjorck et al., 2021), since early stopping alone ensured stable and reproducible generalization.

---

[6] https://pytorch.org/docs/stable/notes/mps.html

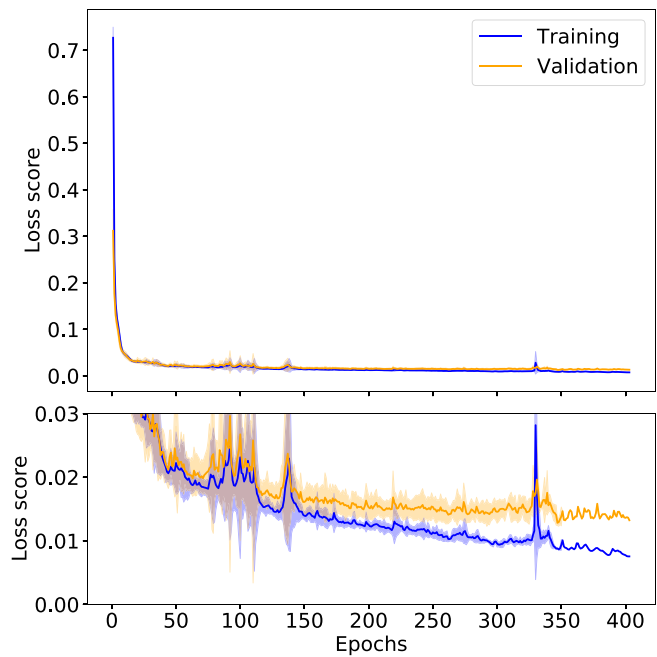

**Fig. 7.** Training and validation loss scores over the epochs of the training process for the MLP architecture with $n = 720$ and $L = 4$, in the case of $I = T = 300$. Solid lines represent the mean loss values across 10 repetitions of the training process, while the shaded regions indicate one standard deviation. The upper panel shows the full loss range, whereas the lower panel provides a zoomed-in view of the 0–0.03 range.

## 5. Results: Model forecasting performance

In this section, we present the performance of the NBM, developed for the ASTRI-Horn monitoring time series using an MLP architecture with 4 hidden layers and 720 neurons per layer, and trained on 300 input and target samples for 112 epochs.

### 5.1. Visual comparison

Out of the 2537 intervals generated using a window size of $W = 600$, we selected four examples from the 254 intervals in the test set for visual comparison. These examples were chosen to be representative of the model's overall predictive performance and to be approximately uniformly distributed across the analyzed period, from September 2022 to July 2024. The corresponding predictions are shown in Figs. 8 and 9, which refer to features related to the Azimuth and Elevation motors, respectively. Within each interval, the first 300 samples represent the model input (solid blue line), while the remaining 300 are used to evaluate the predictions, with actual and predicted values shown as blue and orange dashed lines, respectively. Units are omitted, as all data were normalized through Eq. (7).

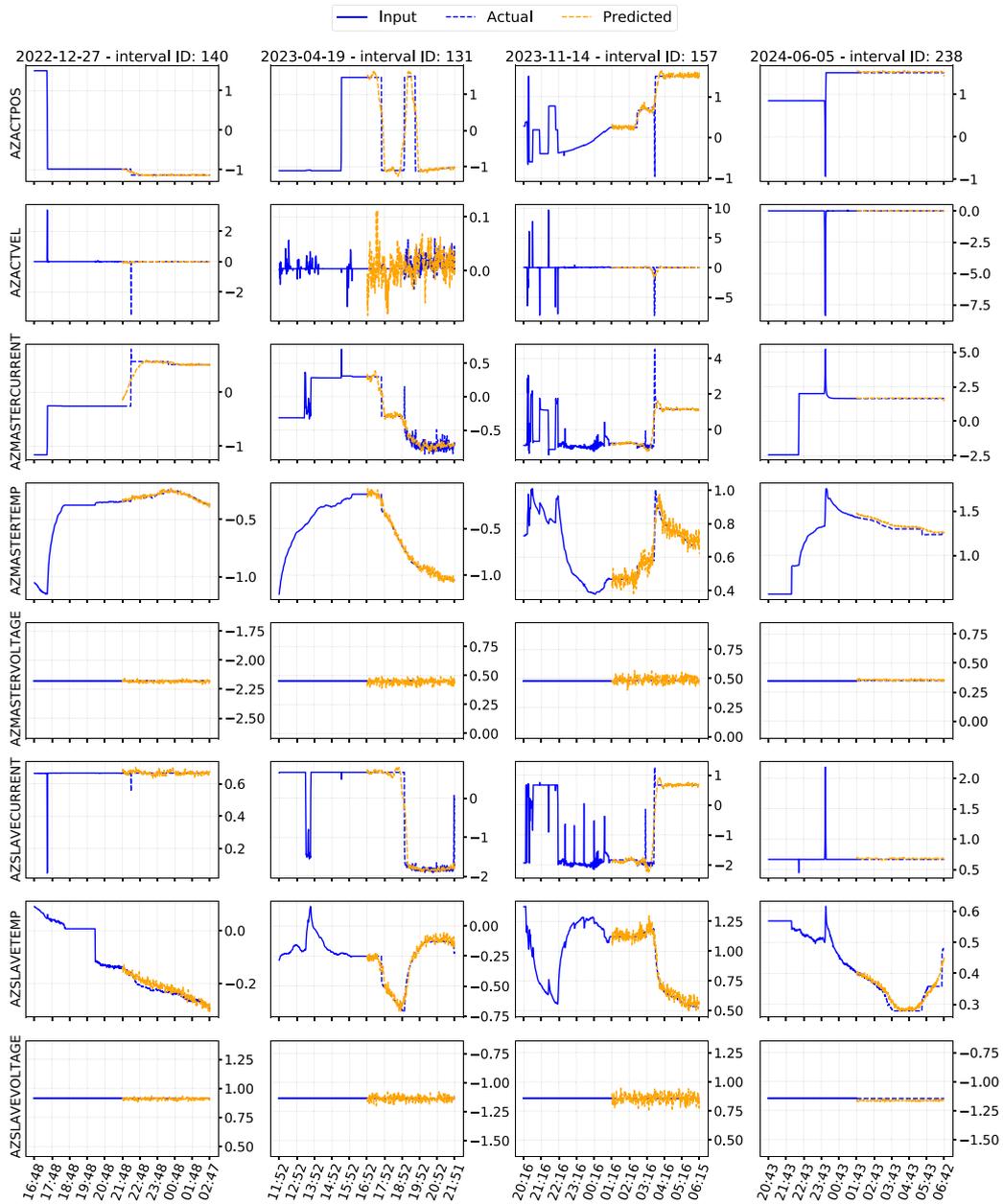

**Fig. 8.** Comparison between actual (dashed blue) and predicted (dashed orange) target samples ($T = 300$) for the features related to the Azimuth motor. The solid blue lines represent the $I = 300$ input samples. Each column denotes a specific interval identified by date and ID. Units are omitted as the data were normalized using z-score (see Section 3.4).

*5.2. MSE and NMAD scores*

To quantitatively assess the model's performance under the selected configuration, we computed both the MSE and NMAD scores (Section 4.2) on the training, validation, and test sets. To account for statistical variability and assess the robustness of the results, we repeated the evaluation over 100 independent trainings using random train/validation/test splits. For each dataset, the median values of MSE and NMAD were computed and are reported in Table 5, along with their associated dispersion, estimated using the NMAD metric.[7] The empirical distributions of these scores for the test set are shown in Fig. 10.

The values of the MSE for the training and test sets are higher than those reported in Table 4 for the same configuration trained for 500 epochs. Although the relative increase is noticeable, the absolute values remain low, making this setting an optimal trade-off between performance and overfitting mitigation. This conclusion is further supported by the NMAD scores, which remain comparably low and consistent across training, validation, and test sets, suggesting good generalization capability.

---

[7] In this case, in Eq. (9)—assuming that $M_i$ denotes the collection of scores obtained for a given metric (either MSE or NMAD)—we set $y_i = M_i$ and $\hat{y}_i = \text{median}(M_i)$.

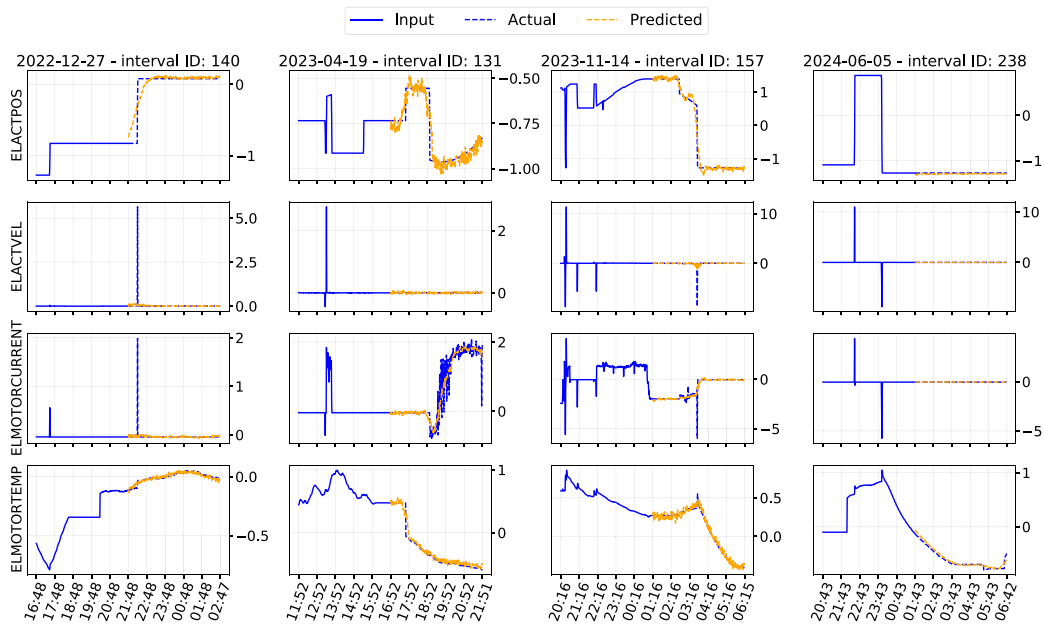

**Fig. 9.** Comparison between actual (dashed blue) and predicted (dashed orange) target samples ($T = 300$) for the features related to the Elevation motor. The solid blue lines represent the $I = 300$ input samples. Each column denotes a specific interval identified by date and ID. Units are omitted as the data were normalized using z-score (see Section 3.4).

**Table 5**
Median MSE and NMAD scores over 100 independent runs for each dataset split. Dispersion is estimated through NMAD.

| Set | MSE | NMAD |
| --- | --- | --- |
| Training | 0.015 ± 0.002 | 0.031 ± 0.009 |
| Validation | 0.018 ± 0.003 | 0.032 ± 0.009 |
| Test | 0.019 ± 0.003 | 0.032 ± 0.009 |

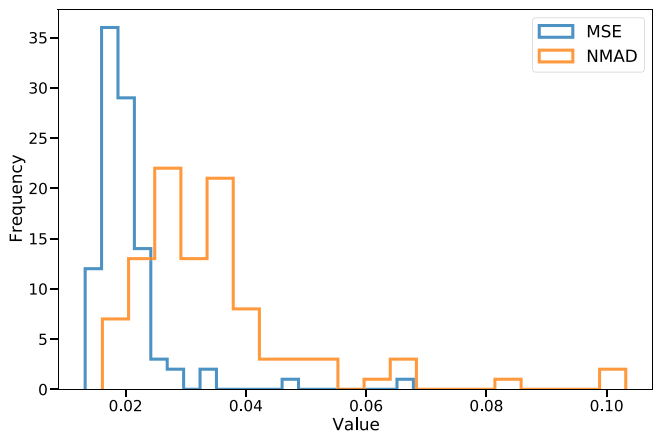

**Fig. 10.** Distributions of MSE and NMAD scores on the test set for 100 independent runs.

The stability of the MSE and NMAD distributions across multiple independent training runs (Fig. 10) suggests that the reported performance is not driven by specific data partitions. No anomalous clusters or unusually low values are observed, providing no empirical evidence of data leakage from the validation or test sets into the training process.

To gain a thorough understanding of the obtained scores, we evaluated these metrics on a per-feature basis. Fig. 11 illustrates the results of this analysis.

For the AZACTVEL, ELACTVEL, and ELMOTORCURRENT variables, the MSE is markedly higher than for the other features, while the corresponding NMAD remains relatively low. This indicates that the typical prediction error is generally small, but that there are a number of large deviations—likely associated with rapid transients or regime changes—that dominate the quadratic metric. In these cases, the model captures the nominal behavior effectively but can occasionally struggle under highly dynamic conditions.

Conversely, for temperatures and voltages, the discrepancy tends to invert, with comparatively smaller MSE values but relatively larger NMAD. This suggests that large deviations are less prominent, while the errors are more uniformly distributed over time, pointing to a reduced responsiveness of the model in tracking finer-scale variations for these features. A similar, though less pronounced, pattern is observed for position-related features.

### 5.3. Residuals

To gain deeper insight into the results presented in Section 5.2, we computed the residuals (Eq. (1)) of the target sequences for each feature in a single run. Fig. 12 plots per-feature residuals, with a dashed orange diagonal indicating perfect predictions. Units are omitted due to the z-score normalization applied to the dataset.

In general, all the features exhibit a strong clustering of points around the diagonal, indicating that the model achieves accurate predictions for the majority of data points. However, the degree of alignment with the diagonal and the spread of the residuals vary across features, reflecting differences in the model's performance for individual features, as already observed in Fig. 11.

Time series related to position, velocity, and current present diverse behaviors depending on the telescope's operational mode: they can vary rapidly during pointing phases, more slowly when tracking, or settle around fixed values if the telescope remains stationary (e.g., for calibration or while awaiting instructions). The vertical strips observed in the residual plots reveal that the model sometimes tends to predict continued motion even when the telescope is effectively stationary, leading to a small fraction of erroneous predictions.

AZACTVEL and ELACTVEL have most of their values concentrated around zero, as the tracking speed is typically on the order of a few arcseconds per second ($\approx 10^{-3}\,°\mathrm{s}^{-1}$), while the stationary speed is zero. The presence of significantly larger values aligning along the zero-prediction line suggests that the model frequently predicts the most probable value rather than capturing large-amplitude variations, which can occur, for example, during rapid telescope repositioning. The same effect extends to other horizontal alignments in the residual plots, notably those observed in AZMASTERCURRENT and ELMOTORCURRENT, highlighting the model's difficulty in forecasting sharp spikes.

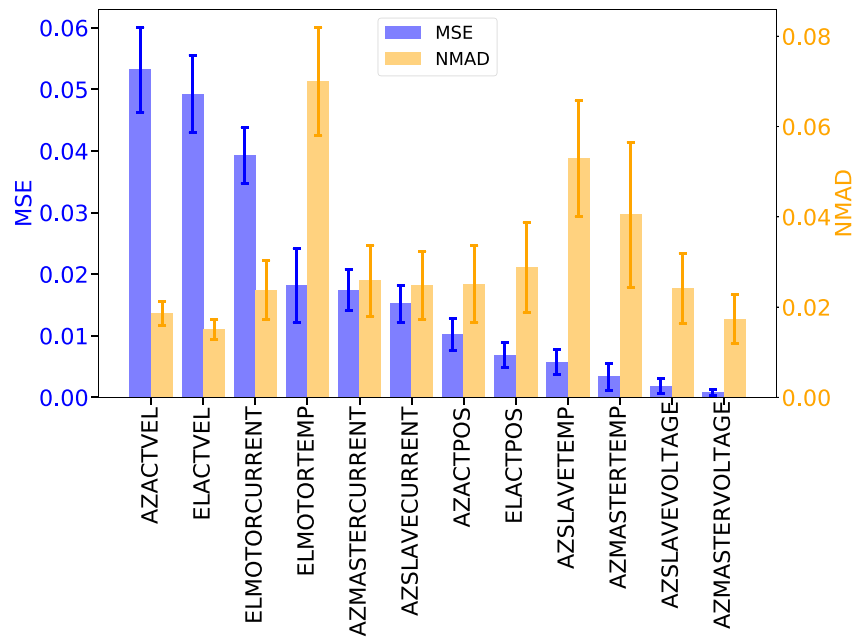

**Fig. 11.** Per-feature evaluation of the MSE and NMAD scores obtained for the MLP model with 7 layers and 720 hidden units per layer, in the case of $I = T = 300$ trained for 112 epochs. Error bars indicate the dispersion over 100 independent runs, estimated via NMAD.

A vertical-strip effect is also observed for AZMASTERVOLTAGE and AZSLAVEVOLTAGE, which exhibit a bimodal distribution (see Fig. 3) and display highly stable signals within each interval (see also Fig. 8). This behavior can be attributed to the fact that, in some cases, when the signals are extremely constant, the model's predictions fluctuate around those levels instead, highlighting its difficulty in forecasting constant signals (see, e.g., interval ID 157 in Fig. 8).

In cases where vertical and horizontal patterns are more prominent, they contribute more to the MSE value. However, as shown in Fig. 11, their impact on the overall predictive performance remains limited. Moreover, since the primary objective is to capture general trends rather than short-term fluctuations, their presence does not undermine the model's effectiveness. On the contrary, the more a feature exhibits a consistently dispersed scatter around the diagonal, the more it contributes to the NMAD, indicating increased difficulty in estimating the typical signal level and, consequently, in reconstructing long-term trends. Nevertheless, although the overall NMAD scores remain consistently low across all features, the residual analysis presented here, together with the results discussed in Section 4.2, allows us to identify the most critical features, informing potential feature-specific refinements during the model production phase.

## 6. Discussion

This study presented a multivariate forecasting model for the ASTRI-Horn prototype's motor-related monitoring data, based on an MLP neural network capable of predicting unseen time series for up to 6.5 h under NOCs using 5 h of input data. The model demonstrated robust predictive performance and strong generalization, achieving an MSE of $0.019 \pm 0.003$ and an NMAD of $0.032 \pm 0.009$ on the test set under its best architecture (4 hidden layers, 720 units per layer, input/forecast lengths of 300 samples each, trained for 112 epochs).

The study specifically focused on 15 features collected by the TCU and related to the Azimuth and Elevation motors. The pre-processing pipeline was essential to the model's success: features were normalized via z-score scaling, data were resampled to 1-minute intervals, and missing values were interpolated to ensure continuity. To focus the analysis on relevant operational periods, the dataset was restricted to active motor intervals, defined by motor status indicators above zero. Correlation analysis guided the removal of three features related to motor torques, which exhibited strong correlations with currents. A sliding-window approach created subsequences of fixed length $W$ from each active interval, increasing the dataset size. Within each window, $I$ samples were used as input for the model, which was tasked with predicting the remaining $T$ samples.

The analysis benchmarked the performance of MLPs across different combinations of $W$, $I$, and $T$, comparing them against LSTMs, which served as a reference due to their ability to model long-term dependencies in time-series data. For $W = 6$, both models performed similarly, exhibiting suboptimal forecasting capacity. At $W = 60$, LSTMs consistently outperformed MLPs, but at $W = 360$, the performance gap narrowed, and for $I = 300$ and $T = 60$, MLPs slightly surpassed LSTMs, suggesting they benefit more from larger datasets. While MLPs require deeper architectures to achieve competitive results, they converge about ten times faster than LSTMs, making them a computationally attractive choice for long-horizon forecasting.

With a fixed $I = 300$, increasing the forecast horizon from 1 h ($T = 60$) to 6.5 h ($T = 390$) led to a pronounced reduction in MSE, plateauing around $T = 300$. At this point, the MLP model with $L = 4$ and $n = 720$ achieved its lowest MSE scores, confirming that this configuration balances complexity and generalization. These findings underscore the model's suitability for multi-hour trend monitoring, making it particularly valuable for extended predictive windows.

The loss scores of this model, evaluated on both training and validation datasets, showed a steady decrease during the early epochs of the training phase, demonstrating effective learning. Faint signs of overfitting emerged after 112 epochs, and training was therefore halted at that point to balance minimal overfitting with model performance.

Under this configuration the model was evaluated over 100 independent runs with random train/validation/test splits, achieving a median MSE of $0.019 \pm 0.003$ and a median NMAD of $0.032 \pm 0.009$ on the test set.

The per-feature MSE and NMAD analysis reveals that AZACT-VEL, ELACTVEL, and ELMOTORCURRENT exhibit higher MSE values but low NMAD, indicating generally accurate predictions with sporadic large deviations, and effective estimation of long-term patterns. Conversely, temperature and voltage features show lower MSE but relatively higher NMAD, suggesting more uniformly distributed errors over time and a slightly reduced sensitivity in capturing fine-scale variations.

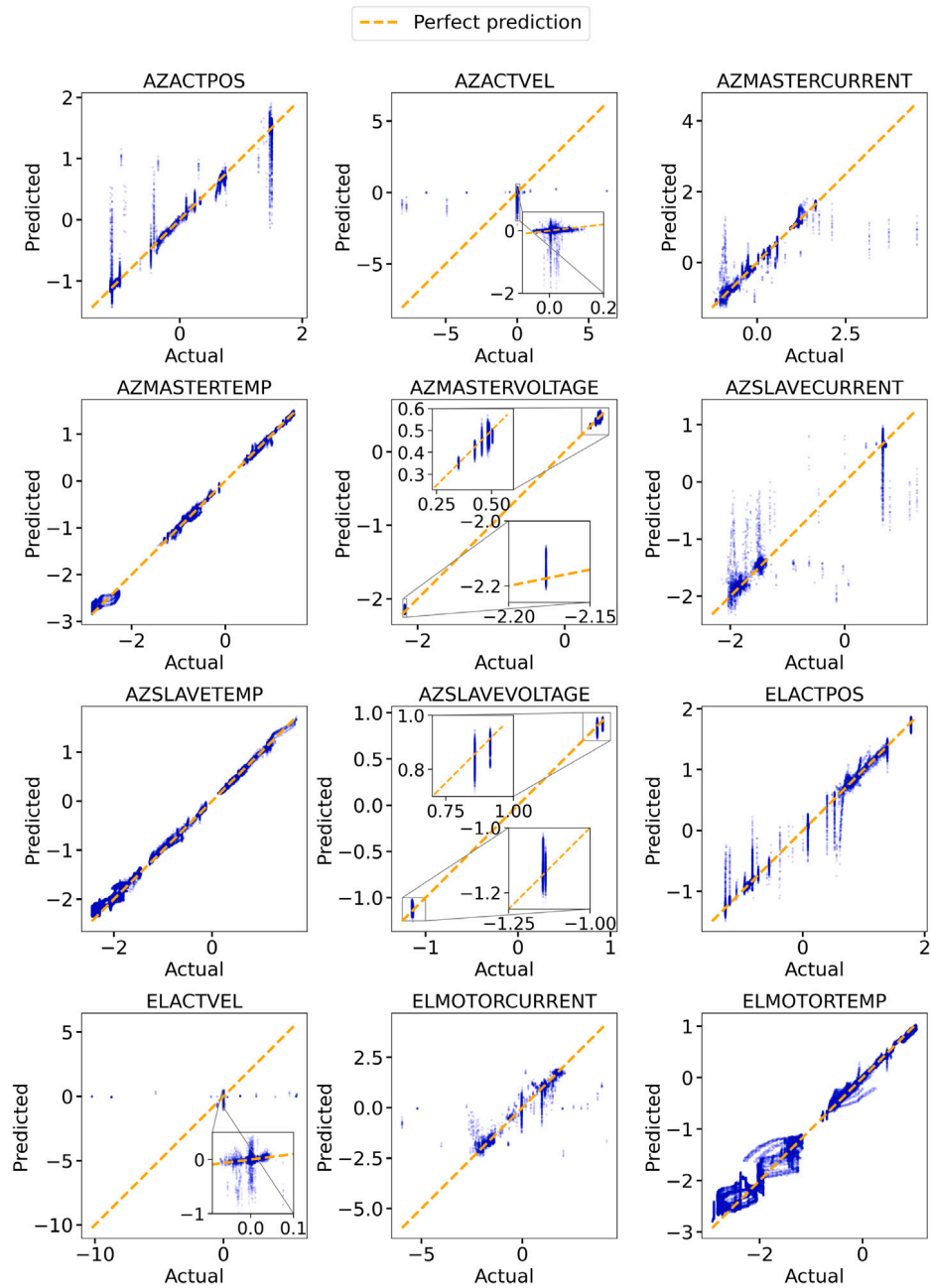

**Fig. 12.** Scatter plot of predicted (*y*-axis) versus actual (*x*-axis) values for each feature for a single run, with the dashed orange diagonal representing perfect predictions. Insets highlight regions of interest. Data are normalized using z-score normalization (see Section 3.4).

Residuals, observed by plotting actual values against predicted ones, exhibited a generally good match, with predicted values closely aligning with the actual ones along the expected diagonal. Minor systematic effects appear as vertical strips in the residual plots, indicating difficulties in forecasting stationary states of the telescope, where the model often predicts motion instead. Similarly, for steady intervals, predictions sometimes fluctuate, revealing challenges in capturing constant signals. The model also struggles to predict spikes, instead defaulting to the most probable value. However, since the main goal is to capture trends rather than short-term fluctuations, these limitations are expected and do not affect the approach's effectiveness. Features exhibiting a consistently dispersed distribution of residuals around the diagonal indicate increased difficulty in estimating the typical signal level and reconstructing long-term trends. These limitations can be mitigated through feature-specific refinement strategies during the model development phase.

It is worth noting that, since ASTRI-Horn was operated as a prototype, no explicit schedule information is available in a time-aligned form. As a consequence, the model relies solely on the dynamical patterns present in the sensor data to infer upcoming regime changes. In future operational configurations, these variables could be incorporated as inputs to improve forecasting accuracy.

The developed NBM model could provide a powerful tool for detecting anomalous deviations from normal behavior up to 6.5 h before they actually manifest. The results obtained suggest that this prediction window could be extended even further, as increasing the amount of training data may, in principle, allow for longer forecasting horizons. Moreover, as the data used in this study included testing phases of the telescope—given its original conception as a prototype—applying this approach to datasets with more extensive operational histories, including several years of scientific observations, could offer valuable opportunities to enhance the model's performance and reliability. In principle, depending on the quantity and quality of the available data, the predictive capability of such a model for a given system could be extended to several days.

This potential is particularly relevant for telescopes located in remote sites, where achieving a prediction window longer than—or at least comparable to—the Mean Time To Repair (MTTR) would be ideal, as it would enable timely maintenance interventions while minimizing downtime. Even if the prediction window were shorter than the MTTR, the model could still provide valuable support in preventing failures or abrupt system interruptions, for instance, by triggering automated safety procedures such as a controlled shutdown upon the detection of anomalous patterns.

## 7. Conclusion

This work introduced a multivariate forecasting model based on a Multi-Layer Perceptron architecture, designed to replicate the behavior of motor-related time series from the ASTRI-Horn Cherenkov telescope under normal operating conditions. The proposed normal behavior model demonstrated robust generalization and effective forecasting capabilities up to 6.5 h. It could provide a powerful tool for enabling early anomaly detection in online monitoring data by comparing real-time measurements with model forecasts to identify statistically significant deviations. In this perspective, the proposed model offers a valuable foundation for the future development of a prognostics and health management system supporting predictive maintenance strategies aimed at mitigating emerging issues before they escalate into critical failures—thus facilitating informed decision-making and reducing unplanned downtime.

Although tailored to ASTRI-Horn, the model is potentially adaptable to other Cherenkov telescopes and arrays—such as the ASTRI Mini-Array of nine telescopes (Scuderi et al., 2022) or the Small-Sized Telescopes of CTAO (Tagliaferri et al., 2022)—as well as to other astronomical infrastructures. Furthermore, while this study specifically focused on the subset of features related to the motors, future and more complex NBMs could be extended to include additional critical mechanical components of astronomical observatories, such as mirror actuators, cooling systems, power components, central actuators, etc.

Future developments will also focus on extending the model to include longer and more representative operational datasets, especially from extended periods of scientific operation, with the goal of further expanding the predictive horizon. In addition, as larger and more diverse datasets become available (e.g., from longer operational histories and/or multiple telescopes), we plan to extend the benchmarking to more complex ANN architectures, such as transformer-based models, to evaluate whether the potential accuracy gains justify the higher computational cost.

Currently, the NBM is intended to be trained offline and then deployed as a static predictor. Possible updates of the model are foreseen as offline retraining or fine-tuning steps when new verified NOC data become available, rather than through continuous online adaptation. Extending the NBM toward an online-learning or reinforcement-learning paradigm represents a possible direction for future investigation, as such approaches could, in principle, enable the model to autonomously capture long-term variations.

A further extension of this work will be to evaluate the model in the presence of fault occurrences. A reliable fault benchmark cannot be established from the current dataset alone, as our previous analysis of the ASTRI-Horn historical monitoring data did not reveal clear or persistent anomaly signatures associated with the few documented failures (Incardona et al., 2024b). One viable approach could involve generating synthetic data to simulate component degradation, which would first require the collection and analysis of at least a limited set of real fault telemetry from operational Cherenkov telescopes. Such empirical data would be essential to characterize how faults manifest in monitoring signals and to guide the development of realistic simulation scenarios.

**CRediT authorship contribution statement**

**F. Incardona:** Writing – review & editing, Writing – original draft, Visualization, Validation, Supervision, Software, Project administration, Methodology, Investigation, Funding acquisition, Formal analysis, Data curation. **A. Costa:** Writing – review & editing, Supervision, Resources, Investigation, Funding acquisition, Conceptualization. **F. Farsian:** Writing – review & editing, Validation, Software, Methodology, Formal analysis, Data curation. **F. Franchina:** Writing – review & editing, Validation, Software, Methodology, Formal analysis, Data curation. **G. Leto:** Writing – review & editing, Supervision, Resources, Investigation, Data curation. **E. Mastriani:** Writing – review & editing, Conceptualization. **K. Munari:** Writing – review & editing, Conceptualization. **G. Pareschi:** Supervision, Resources. **S. Scuderi:** Resources. **S. Spinello:** Writing – review & editing, Conceptualization. **G. Tosti:** Resources.

**Declaration of Generative AI and AI-assisted technologies in the writing process**

During the preparation of this work the authors used ChatGPT in order to improve readability and language. After using this tool, the authors reviewed and edited the content as needed and take full responsibility for the content of the published article.

**Funding sources**

This work was supported by the "Istituto Nazionale di Astrofisica", Italy (INAF) Mini-Grant, and by the Italian Research Center on High Performance Computing, Big Data and Quantum Computing (ICSC), a project funded by the European Union—NextGenerationEU—and the National Recovery and Resilience Plan (NRRP)—Mission 4 Component 2—within the activities of Spoke 2 (Fundamental Research and Space Economy).

**Declaration of competing interest**

The authors declare that they have no known competing financial interests or personal relationships that could have appeared to influence the work reported in this paper.

**Acknowledgments**

This work was conducted in the context of the ASTRI Project. We gratefully acknowledge support from the people, agencies, and organizations listed here: http://www.astri.inaf.it/en/library/. This paper went through the internal ASTRI review process.

## References

Abdulrahim, H., Alshibani, S.M., Ibrahim, O., Elhag, A.A., 2025. Prediction OPEC oil price utilizing long short-term memory and multi-layer perceptron models. Alex. Eng. J. 110, 607–612. doi:10.1016/j.aej.2024.10.057.

Aggarwal, C.C., 2017. Outlier Analysis, second ed. Springer, doi:10.1007/978-3-319-47578-3.

Ali, A., Abdelhadi, A., 2022. Condition-based monitoring and maintenance: State of the art review. Appl. Sci. 12 (2), doi:10.3390/app12020688.

Arshad, K., Ali, R.F., Muneer, A., Aziz, I.A., Naseer, S., Khan, N.S., Taib, S.M., 2022. Deep reinforcement learning for anomaly detection: A systematic review. IEEE Access 10, 124017–124035. doi:10.1109/ACCESS.2022.3224023.

Bjorck, J., Weinberger, K.Q., Gomes, C., 2021. Understanding Decoupled and Early Weight Decay, vol. 35, pp. 6777–6785. doi:10.1609/aaai.v35i8.16837.

Blázquez-García, A., Conde, A., Mori, U., Lozano, J.A., 2021. A review on outlier/anomaly detection in time series data. ACM Comput. Surv. 54 (3), 56:1–56:33. doi:10.1145/3444690.

Boggia, L., de Lima, R.T., Malaescu, B., 2025. Benchmarking unsupervised strategies for anomaly detection in multivariate time series. doi:10.48550/arXiv.2506.20574, arXiv:2506.20574.

Bulgarelli, A., Lucarelli, F., Tosti, G., Conforti, V., Parmiggiani, N., Schwarz, J.H., Gallardo, J.G.A., Antonelli, L.A., Araya, M., Balbo, M., Baroncelli, L., Bigongiari, C., Bruno, P., Capalbi, M., Cardillo, M., Castillo, G.A.R., Catalano, O., Compagnino, A.A., Corpora, M., Costa, A., Crestan, F., Cusumano, G., D'Aì, A., Fioretti, V., Gallozzi, S., Germani, S., Gianotti, F., Giordano, V., Giuliani, A., Grillo, A., Huerta, I., Incardona, F., Iovenitti, S., Palombara, N.L., Parola, V.L., Landoni, M., Lombardi, S., Maccarone, M.C., Millul, R., Mineo, T., Montenegro, G., Mollica, D., Munari, K., Pagliaro, A., Pareschi, G., Pastore, V., Perri, M., Pintore, F., Romano, P., Russo, F., Sanchez, R.Z., Sangiorgi, P., Saturni, F.G., Sayes, N., Sciacca, E., Sliusar, V., Scuderi, S., Tacchini, A., Testa, V., Trifoglio, M., Tutone, A., Vercellone, S., Walter, R., for the ASTRI Project, 2024. Software architecture and development approach for the ASTRI Mini-Array project at the Teide Observatory. J. Astron. Telesc. Instruments, Syst. 10 (1), 017001. doi:10.1117/1.JATIS.10.1.017001.

Canestrari, R., Cascone, E., Conconi, P., Fiorini, M., Giro, E., Palombara, N.L., Lessio, L., Pareschi, G., Rodeghiero, G., Sironi, G., Stringhetti, L., Toso, G., Tosti, G., Martelli, F., Parodi, G., Rossettini, P., Tomelleri, R., 2013. The ASTRI SST-2M prototype for the next generation of Cherenkov telescopes: structure and mirrors. In: O'Dell, S.L., Pareschi, G. (Eds.), In: Optics for EUV, X-Ray, and Gamma-Ray Astronomy VI, vol. 8861, SPIE, International Society for Optics and Photonics, 886102. doi:10.1117/12.2024379.

Canestrari, R., Gargano, C., Sottile, G., Biondo, B., Bonanno, G., Bruno, P., Capalbi, M., Catalano, O., Conforti, V., Garozzo, S., Gianotti, F., Giarrusso, S., Giro, E., Grillo, A., Impiombato, D., Rosa, G.L., Maccarone, M.C., Marano, D., Mineo, T., Pareschi, G., Romeo, G., Russo, F., Sangiorgi, P., Scuderi, S., Segreto, A., Sironi, G., 2019. The innovative Cherenkov camera based on SiPM sensors of the ASTRI-Horn telescope: from the T/M and electrical design to the full assembly and testing in a harsh environment. In: James, R.B., Burger, A., Payne, S.A. (Eds.), In: Hard X-Ray, Gamma-Ray, and Neutron Detector Physics XXI, vol. 11114, SPIE, International Society for Optics and Photonics, p. 111140A. doi:10.1117/12.2528153.

Catalano, O., Capalbi, M., Gargano, C., Giarrusso, S., Impiombato, D., Rosa, G.L., Maccarone, M.C., Mineo, T., Russo, F., Sangiorgi, P., Segreto, A., Sottile, G., Biondo, B., Bonanno, G., Garozzo, S., Grillo, A., Marano, D., Romeo, G., Scuderi, S., Canestrari, R., Conconi, P., Giro, E., Pareschi, G., Sironi, G., Conforti, V., Gianotti, F., Gimenes, R., 2018. The ASTRI camera for the Cherenkov Telescope Array. In: Evans, C.J., Simard, L., Takami, H. (Eds.), In: Ground-Based and Airborne Instrumentation for Astronomy VII, vol. 10702, SPIE, International Society for Optics and Photonics, 1070237. doi:10.1117/12.2314984.

Chesterman, X., Verstraeten, T., Daems, P.-J., Nowé, A., Helsen, J., 2023. Overview of normal behavior modeling approaches for SCADA-based wind turbine condition monitoring demonstrated on data from operational wind farms. Wind. Energy Sci. 8 (6), 893–924. doi:10.5194/wes-8-893-2023.

Conforti, V., Gianotti, F., Pastore, V., Trifoglio, M., Bulgarelli, A., Addis, A., Baroncelli, L., Capalbi, M., Fioretti, V., Parmiggiani, N., Sangiorgi, P., Corpora, M., Catalano, O., Costa, A., Incardona, F., Russo, F., 2022. The Array Data Acquisition System software architecture of the ASTRI Mini-Array Project. Submitt. Soc. Photo-Optical Instrum. Eng. (SPIE) Conf. Ser. doi:10.1117/12.2626600.

Conforti, V., Trifoglio, M., Gianotti, F., Malaguti, G., Bulgarelli, A., Fioretti, V., Zoli, A., Catalano, O., Capalbi, M., Sangiorgi, P., 2016. Software design of the ASTRI camera server proposed for the Cherenkov Telescope Array. In: Chiozzi, G., Guzman, J.C. (Eds.), In: Software and Cyberinfrastructure for Astronomy IV, vol. 9913, SPIE, International Society for Optics and Photonics, p. 99133G. doi:10.1117/12.2230016.

Correia, L., Goos, J., Klein, P., Bäck, T., Kononova, A.V., 2024. Online model-based anomaly detection in multivariate time series: Taxonomy, survey, research challenges and future directions. Eng. Appl. Artif. Intell. 138, 109323. doi:10.1016/j.engappai.2024.109323.

Costa, A., Mastriani, E., Incardona, F., Munari, K., Spinello, S., 2024. Predictive maintenance study for high-pressure industrial compressors: Hybrid clustering models. doi:10.48550/arXiv.2411.13919, arXiv:2411.13919.

Costa, A., Munari, K., Incardona, F., et al., 2021. The Monitoring, Logging, and Alarm system for the Cherenkov Telescope Array. In: Proceedings of 37th International Cosmic Ray Conference — PoS, Vol. 395. ICRC2021, p. 700. doi:10.22323/1.395.0700.

Costa, A., Tosti, G., Schwarz, J., Bruno, P., Bulgarelli, A., Calanducci, A., Conforti, V., Germani, S., Gianotti, F., Grillo, A., Incardona, F., Munari, K., Russo, F., Sciacca, E., Vitello, F., 2020. Architectural design and prototype for the logging, monitoring, and alarm system for the ASTRI mini-array. In: Guzman, J.C., Ibsen, J. (Eds.), Software and Cyberinfrastructure for Astronomy VI. vol. 11452, SPIE, International Society for Optics and Photonics, doi:10.1117/12.2560697.

Dubey, S.R., Singh, S.K., Chaudhuri, B.B., 2022. Activation functions in deep learning: A comprehensive survey and benchmark. Neurocomputing 503, 92–108. doi:10.1016/j.neucom.2022.06.111.

Gambadoro, S., et al., 2023. Predictive maintenance for array of Cherenkov telescopes. In: Machine Learning for Astrophysics. Springer International Publishing, Cham, pp. 205–208. doi:10.1007/978-3-031-34167-0_41.

Giro, E., Canestrari, R., Sironi, G., Antolini, E., Conconi, P., Fermino, C.E., Gargano, C., Rodeghiero, G., Russo, F., Scuderi, S., Tosti, G., Vassiliev, V., Pareschi, G., 2017. First optical validation of a Schwarzschild Couder telescope: the ASTRI SST-2M Cherenkov telescope. Astron. Astrophys. 608, A86. doi:10.1051/0004-6361/201731602.

Hughes, A., Drury, B., 2019. Chapter 5 - Induction motors—Rotating field, slip and torque. In: Hughes, A., Drury, B. (Eds.), Electric Motors and Drives (Fifth Edition), fifth ed. Newnes, pp. 161–190. doi:10.1016/B978-0-08-102615-1.00005-2.

Incardona, F., Costa, A., Leto, G., Munari, K., Pareschi, G., Scuderi, S., Tosti, G., 2024a. Revealing predictive maintenance strategies from comprehensive data analysis of ASTRI-horn historical monitoring data. In: Submitted To Astronomical Data Analysis Software and Systems (ADASS) XXXIII Conference Series. doi:10.48550/arXiv.2406.07308, arXiv:2406.07308.

Incardona, F., Costa, A., Leto, G., Munari, K., Pareschi, G., Scuderi, S., Tosti, G., Mastriani, E., Spinello, S., 2024b. Tracing component wear signatures in ASTRI-Horn historical monitoring time-series. In: Ibsen, J., Chiozzi, G. (Eds.), In: Software and Cyberinfrastructure for Astronomy VIII, vol. 13101, SPIE, International Society for Optics and Photonics, 1310122. doi:10.1117/12.3014702.

Incardona, F., Costa, A., Munari, K., Bruno, P., Bulgarelli, A., Germani, S., Grillo, A., Schwarz, J., Sciacca, E., Tosti, G., Vitello, F., Tudisco, G., 2021a. LOgging UnifieD for ASTRI mini array. In: Proceedings of Science, 37th International Cosmic Ray Conference. ICRC2021, p. 195. doi:10.22323/1.395.0195.

Incardona, F., Costa, A., Munari, K., Bruno, P., Germani, S., Grillo, A., Oya, I., Neise, D., Sciacca, E., 2021b. Optimization of the storage database for the Monitoring system of the Cherenkov Telescope Array. In: Astronomical Data Analysis Software and Systems (ADASS) XXXI Conference Series. doi:10.48550/arXiv.2207.06381, Cape Town, South Africa and Online.

Incardona, F., Costa, A., Munari, K., Gambadoro, S., Germani, S., Bruno, P., Bulgarelli, A., Conforti, V., Gianotti, F., Grillo, A., Pastore, V., Russo, D., Schwarz, J., Tosti, G., Cavalieri, S., 2022. The monitoring, logging, and alarm system of the ASTRI mini-array gamma-ray air-Cherenkov experiment at the Observatorio del Teide. Soc. Photo-Optical Instrum. Eng. (SPIE) Conf. Ser. doi:10.1117/12.2629887.

Jawad, S., Jaber, A., 2021. Rolling bearing fault detection based on vibration signal analysis and cumulative sum control chart. FME Trans. 49, doi:10.5937/fme2103684M.

Jia, W., Shukla, R.M., Sengupta, S., 2019. Anomaly detection using supervised learning and multiple statistical methods. In: 2019 18th IEEE International Conference on Machine Learning and Applications. ICMLA, pp. 1291–1297. doi:10.1109/ICMLA.2019.00211.

Johnson, R., Wichern, D., 1988. Applied multivariate statistical analysis. In: Prentice-Hall Series in Statistics, Prentice-Hall.

Katragadda, S., Odubade, K., Isabirye, D., 2020. Anomaly detection detecting unusual behavior using machine learning algorithms to identify potential security threats or system failures. Int. Res. J. Mod. Eng. Technol. Sci. 2, 2582–5208. doi:10.56726/IRJMETS1335.

Khandelwal, P., Konar, J., Brahma, B., 2020. Training RNN and it's variants using sliding window technique. In: 2020 IEEE International Students' Conference on Electrical,Electronics and Computer Science. SCEECS, pp. 1–5. doi:10.1109/SCEECS48394.2020.93.

Kingma, D.P., Ba, J., 2014. Adam: A method for stochastic optimization. doi:10.48550/arXiv.1412.6980, CoRR, abs/1412.6980.

Lazcano, A., Jaramillo-Morán, M.A., Sandubete, J.E., 2024. Back to basics: The power of the multilayer perceptron in financial time series forecasting. Mathematics 12 (12), doi:10.3390/math12121920.

Lei, Y., Li, N., Guo, L., Li, N., Yan, T., Lin, J., 2020. Machinery health prognostics: A systematic review from data acquisition to RUL prediction. Mech. Syst. Signal Process. 104761, doi:10.1016/j.ymssp.2017.11.016.

Leto, G., Bellassai, C., Bigongiari, C., Biondo, B., Bruno, G., Capalbi, M., Catalano, O., Conforti, V., Contino, G., Compagnino, A., Corpora, M., Crestan, S., Gargano, C., Garozzo, S., Gianotti, F., Giordano, V., Grillo, A., Iovenitti, S., La Palombara, N., Tosti, G., 2023. Status and performance of the ASTRI-horn dual mirror air-Cherenkov telescope after a major maintenance and refurbishment intervention. In: 38th International Cosmic Ray Conference. p. 729. doi:10.22323/1.444.0729.

Leto, G., Maccarone, M.C., Bellassai, G., Bruno, P., Fiorini, M., Grillo, A., Martinetti, E., Rosa, G.L., Segreto, A., Sottile, G., Stringhetti, L., 2014. The site of the ASTRI SST-2M telescope prototype: Atmospheric monitoring and auxiliary instrumentation. In: Proceedings of the Atmospheric Monitoring for High-Energy Astroparticle Detectors (AtmoHEAD) Conference. doi:10.48550/arXiv.1402.3515, arXiv:1402.3515.

Li, M., 2024. Comprehensive review of backpropagation neural networks. Acad. J. Sci. Technol. 9 (1), 150–154. doi:10.54097/51y16r47.

Li, G., Jung, J.J., 2023. Deep learning for anomaly detection in multivariate time series: Approaches, applications, and challenges. Inf. Fusion 91, 93–102. doi:10.1016/j.inffus.2022.10.008.

Lima, F., Alves de Souza, F., 2023. A large comparison of normalization methods on time series. Big Data Res. 34, 100407. doi:10.1016/j.bdr.2023.100407.

Lindemann, B., Maschler, B., Sahlab, N., Weyrich, M., 2021. A survey on anomaly detection for technical systems using LSTM networks. Comput. Ind. 131, 103498. doi:10.1016/j.compind.2021.103498.

Lombardi, S., Catalano, O., Scuderi, S., Antonelli, L.A., Pareschi, G., Antolini, E., Arrabito, L., Bellassai, G., Bernlöhr, K., Bigongiari, C., Biondo, B., Bonanno, G., Bonnoli, G., Böttcher, G.M., Bregeon, J., Bruno, P., Canestrari, R., Capalbi, M., Caraveo, P., Conconi, P., Conforti, V., Contino, G., Cusumano, G., de Gouveia Dal Pino, E.M., Distefano, A., Farisato, G., Fermino, C., Fiorini, M., Frigo, A., Gallozzi, S., Gargano, C., Garozzo, S., Gianotti, F., Giarrusso, S., Gimenes, R., Giro, E., Grillo, A., Impiombato, D., Incorvaia, S., La Palombara, N., La Parola, V., La Rosa, G., Leto, G., Lucarelli, F., Maccarone, M.C., Marano, D., Martinetti, E., Miccichè, A., Millul, R., Mineo, T., Nicotra, G., Occhipinti, G., Pagano, I., Perri, F., Romeo, G., Russo, F., Sacco, B., Sangiorgi, P., Saturni, F.G., Segreto, A., Sironi, G., Sottile, G., Stamerra, A., Stringhetti, L., Tagliaferri, G., Tavani, M., Testa, V., Timpanaro, M.C., Toso, G., Tosti, G., Trifoglio, M., Umana, G., Vercellone, S., Zanmar Sanchez, R., Arcaro, C., Bulgarelli, A., Cardillo, M., Cascone, E., Costa, A., D'Aì, A., D'Ammando, F., Del Santo, M., Fioretti, V., Lamastra, A., Mereghetti, S., Pintore, F., Rodeghiero, G., Romano, P., Schwarz, J., Sciacca, E., Vitello, F.R., Wolter, A., 2020. First detection of the Crab Nebula at TeV energies with a Cherenkov telescope in a dual-mirror Schwarzschild-Couder configuration: the ASTRI-horn telescope. Astron. Astrophys. 634, A22. doi:10.1051/0004-6361/201936791.

Luo, Z., Li, Y., Lu, J., Chen, Z., Fu, L., Zhang, S., Xiao, H., Du, W., Gong, Y., Shu, C., Ma, W., Meng, X., Zhou, X., Fan, Z., 2024. Photometric redshift estimation for CSST survey with LSTM neural networks. Mon. Not. R. Astron. Soc. 535 (2), 1844–1855. doi:10.1093/mnras/stae2446.

Maccarone, M.C., Leto, G., Bruno, P., Fiorini, M., Grillo, A., Segreto, A., Stringhetti, L., 2013. The site of the ASTRI SST-2M telescope prototype. In: Proceedings of the 33rd International Cosmic Ray Conference. ICRC2013, doi:10.48550/arXiv.1307.5139, arXiv:1307.5139.

Molęda, M., Małysiak-Mrozek, B., Ding, W., Sunderam, V., Mrozek, D., 2023. From corrective to predictive maintenance—A review of maintenance approaches for the power industry. Sensors 23 (13), doi:10.3390/s23135970.

Ngo Bibinbe, A.M.S., Mbouopda, M.F., Mbiadou Saleu, G.R., Mephu Nguifo, E., 2022. A survey on unsupervised learning algorithms for detecting abnormal points in streaming data. In: 2022 International Joint Conference on Neural Networks. IJCNN, pp. 1–8. doi:10.1109/IJCNN55064.2022.9892195.

Page, E.S., 1954. Continuous inspection schemes. Biometrika 41 (1/2), 100–115. doi:10.2307/2333009.

Pareschi, G., 2016. The ASTRI SST-2M prototype and mini-array for the Cherenkov Telescope Array (CTA). In: Ground-Based and Airborne Telescopes VI. In: Society of Photo-Optical Instrumentation Engineers (SPIE) Conference Series, vol. 9906, p. 99065T. doi:10.1117/12.2232275.

Pastore, V., Conforti, V., Gianotti, F., Bulgarelli, A., Parmiggiani, N., Incardona, F., Costa, A., Russo, F., 2022. Array Data Acquisition System interface for online distribution of acquired data in the ASTRI Mini-Array project. Submitt. Soc. Photo-Optical Instrum. Eng. (SPIE) Conf. Ser. doi:10.1117/12.2629922.

Paszke, A., Gross, S., Massa, F., Lerer, A., Bradbury, J., Chanan, G., Killeen, T., Lin, Z., Gimelshein, N., Antiga, L., Desmaison, A., Kopf, A., Yang, E., DeVito, Z., Raison, M., Tejani, A., Chilamkurthy, S., Steiner, B., Fang, L., Bai, J., Chintala, S., 2019. PyTorch: An imperative style, high-performance deep learning library. In: Advances in Neural Information Processing Systems 32. Curran Associates, Inc., pp. 8024–8035. doi:10.48550/arXiv.1912.01703.

Pugliese, R., Regondi, S., Marini, R., 2021. Machine learning-based approach: global trends, research directions, and regulatory standpoints. Data Sci. Manag. 4, 19–29. doi:10.1016/j.dsm.2021.12.002.

Rodeghiero, G., Giro, E., Canestrari, R., Pernechele, C., Sironi, G., Pareschi, G., Lessio, L., Conconi, P., 2016. Qualification and testing of a large hot slumped secondary mirror for Schwarzschild–Couder imaging air Cherenkov telescopes. Publ. Astron. Soc. Pac. 128 (963), 055001. doi:10.1088/1538-3873/128/963/055001.

Russo, F., Tosti, G., Bruno, P., Bulgarelli, A., Gianotti, F., Parmiggiani, N., Capalbi, M., Conforti, V., Costa, A., Fiori, M., Germani, S., Grillo, A., Sangiorgi, P., Schwarz, J., Scuderi, S., Zampieri, L., Incardona, F., Munari, K., Pastore, V., 2022. The telescope control system for the ASTRI mini-array of imaging atmospheric Cherenkov telescopes. Submitt. Soc. Photo-Optical Instrum. Eng. (SPIE) Conf. Ser. doi:10.1117/12.2629943.

Salehin, I., Kang, D.-K., 2023. A review on dropout regularization approaches for deep neural networks within the scholarly domain. Electronics 12 (14), doi:10.3390/electronics12143106.

Scuderi, S., CTA ASTRI Project Collaboration, 2018. From the Etna volcano to the Chilean Andes: ASTRI end-to-end telescopes for the Cherenkov telescope array. In: Angeli, G.Z., Dierickx, P. (Eds.), Proc. SPIE Int. Soc. Opt. Eng. 10700, 107005Z. doi:10.1117/12.2312453.

Scuderi, S., et al., 2022. The ASTRI mini-array of Cherenkov telescopes at the Observatorio del Teide. J. High Energy Astrophys. 35, 52–68. doi:10.1016/j.jheap.2022.05.001, arXiv:2208.04571.

Sgueglia, A., Di Sorbo, A., Visaggio, C.A., Canfora, G., 2022. A systematic literature review of IoT time series anomaly detection solutions. Future Gener. Comput. Syst. 134, 170–186. doi:10.1016/j.future.2022.04.005.

Sulich, A., Tosti, G., Baldini, V., Coretti, I., Di Marcantonio, P., Pareschi, G., Scuderi, S., 2024. The PLC control system development from ASTRI-Horn to ASTRI Mini-Array. In: Ibsen, J., Chiozzi, G. (Eds.), Software and Cyberinfrastructure for Astronomy VIII. vol. 13101, SPIE, International Society for Optics and Photonics, 1310131. doi:10.1117/12.3018777.

Tagliaferri, G., Antonelli, A., Arnesen, T., Aschersleben, J., Attinà, P., Balbo, M., Bang, S., Barcelo, M., Baryshev, A., Bellassai, G., Berge, D., Bicknell, C., Bigongiari, C., Bonnoli, G., Bouley, F., Brown, A., Bulgarelli, A., Cappi, M., Caraveo, P., Caschera, S., Chadwick, P., Conte, F., Cotter, G., Cristofari, P., De Frondat, F., De Gouveia Dal Pino, E., De Simone, N., Depaoli, P., Dournaux, J.L., Duffy, C., Einecke, S., Fermino, C., Funk, S., Gargano, C., Giavitto, G., Giuliani, A., Greenshaw, T., Hinton, J., Huet, J.M., Iovenitti, S., La Palombara, N., Lapington, J., Laporte, P., Leach, S., Lessio, L., Leto, G., Lloyd, S., Lucarelli, F., Lombardi, S., Macchi, A., Martinetti, E., Miccichè, A., Millul, R., Mineo, T., Mitsunari, T., Nayak, N., Nicotra, G., Okumura, A., Pareschi, G., Penno, M., Prokoph, H., Rebert, E., Righi, C., Rulten, C., Russo, F., Sanchez, R.Z., Saturni, F.G., Schaefer, J., Schwab, B., Scuderi, S., Sironi, G., Sliusar, V., Sol, H., Spencer, A., Stamerra, A., Tajima, H., Tavecchio, F., Tosti, G., Trois, A., Vecchi, M., Vercellone, S., Vink, J., Walter, R., Watson, J., White, R., Zanin, R., Zampieri, L., Zech, A., Zink, A., 2022. The small-sized telescope of CTAO. In: Marshall, H.K., Spyromilio, J., Usuda, T. (Eds.), Ground-Based and Airborne Telescopes IX. In: Society of Photo-Optical Instrumentation Engineers (SPIE) Conference Series, vol. 12182, p. 121820K. doi:10.1117/12.2627956.

Tanci, C., et al., 2016. The ASTRI mini-array software system (MASS) implementation: a proposal for the Cherenkov Telescope Array. In: Software and Cyberinfrastructure for Astronomy IV. In: Society of Photo-Optical Instrumentation Engineers (SPIE) Conference Series, vol. 9913, p. 99133L. doi:10.1117/12.2231294.

Taşcı, B., Omar, A., Ayvaz, S., 2023. Remaining useful lifetime prediction for predictive maintenance in manufacturing. Comput. Ind. Eng. 184, 109566. doi:10.1016/j.cie.2023.109566.

The CTA Consortium, 2018. Science with the Cherenkov Telescope Array. WORLD SCIENTIFIC, doi:10.1142/10986.

Tong, X., Jung, W., Banning, J., 2023. A fine-grained semi-supervised anomaly detection framework for predictive maintenance of industrial assets. Annu. Conf. the PHM Soc. 15, doi:10.36001/phmconf.2023.v15i1.3547.

van Engelen, J.E., Hoos, H.H., 2020. A survey on semi-supervised learning. Mach. Learn. 109 (2), 373–440. doi:10.1007/s10994-019-05855-6.

Vercellone, S., et al., ASTRI Collaboration, 2013. The ASTRI Mini-Array Science Case. In: 33rd International Cosmic Ray Conference. p. 0109. doi:10.48550/arXiv.1307.5671, arXiv:1307.5671.

Vercellone, S., et al., 2022. ASTRI mini-array core science at the Observatorio del Teide. J. High Energy Astrophys. 35, 1–42. doi:10.1016/j.jheap.2022.05.005, arXiv:2208.03177.

Xiao, X., Liu, J., Liu, D., Tang, Y., Qin, S., Zhang, F., 2022. A normal behavior-based condition monitoring method for wind turbine main bearing using dual attention mechanism and Bi-LSTM. Energies 15 (22), doi:10.3390/en15228462.

Xie, L., Zou, S., Xie, Y., Veeravalli, V.V., 2021. Sequential (Quickest) change detection: Classical results and new directions. IEEE J. Sel. Areas Inf. Theory 2 (2), 494–514. doi:10.1109/JSAIT.2021.3072962.

Yokkampon, U., Chumkamon, S., Mowshowitz, A., Fujisawa, R., Hayashi, E., 2021. Anomaly detection using support vector machines for time series data. J. Robot. Netw. Artif. Life 8 (1), 41–46. doi:10.2991/jrnal.k.210521.010.

Zamanzadeh Darban, Z., Webb, G.I., Pan, S., Aggarwal, C.C., Salehi, M., 2024. Deep learning for time series anomaly detection: A survey. ACM Comput. Surv. 57 (1), 1–42. doi:10.1145/3691338.

Zio, E., 2022. Prognostics and health management (PHM): Where are we and where do we (need to) go in theory and practice. Reliab. Eng. Syst. Saf. 218, 108119. doi:10.1016/j.ress.2021.108119.